\documentclass[runningheads]{llncs}
\newcommand{\repeatthanks}{\textsuperscript{\thefootnote}}
\usepackage{color,soul}
\usepackage{graphicx}
\usepackage{multirow}
\usepackage{url}
\usepackage[english]{babel}
\usepackage[utf8]{inputenc}
\usepackage{bbding}
\makeatletter
\newcommand{\printfnsymbol}[1]{%
  \textsuperscript{\@fnsymbol{#1}}%
}
\makeatother

\begin{document}

\title{Automatic Segmentation of Muscle Tissue and Inter-muscular Fat in Thigh and Calf MRI Images}
\titlerunning{Automatic Segmentation of Muscle Tissue and Inter-muscular Fat}

\author{Rula Amer\inst{1}$^{,(}$\Envelope$^)$\orcidID{0000-0002-5630-827X} \and
Jannette Nassar\inst{1}\orcidID{0000-0002-4936-7298} \and David Bendahan\inst{2} \and Hayit Greenspan\inst{1}$^,$\thanks{Equal contribution}\orcidID{0000-0001-6908-7552} \and
Noam Ben-Eliezer\inst{1,3,4}$^,$\repeatthanks\orcidID{0000-0003-2944-6412}}
%
%

\institute{Department of Biomedical Engineering, Tel Aviv University, Tel Aviv, Israel \email{rolaamer@mail.tau.ac.il} \and
Aix Marseille Univ, CNRS, CRMBM, Marseille, France 
\and Center for Advanced
Imaging Innovation and Research, New York University, New York, NY, United States \and  Sagol School of Neuroscience, Tel Aviv University, Tel Aviv, Israel
} 
\authorrunning{R. Amer et al.}
%
%
\maketitle  
\begin{abstract}
Magnetic resonance imaging (MRI) of thigh and calf muscles is one of the most effective techniques for estimating fat infiltration into muscular dystrophies. The infiltration of adipose tissue into the diseased muscle region varies in its severity across, and within, patients. In order to efficiently quantify the infiltration of fat, accurate segmentation of muscle and fat is needed. An estimation of the amount of infiltrated fat is typically done visually by experts. Several algorithmic solutions have been proposed for automatic segmentation. While these methods may work well in mild cases, they struggle in moderate and severe cases due to the high variability in the intensity of infiltration, and the tissue's heterogeneous nature. To address these challenges, we propose a deep-learning approach, producing robust results with high Dice Similarity Coefficient (DSC) of $0.964$, $0.917$ and $0.933$ for muscle-region, healthy muscle and inter-muscular adipose tissue (IMAT) segmentation, respectively.

\end{abstract}
\section{Introduction}

Muscle dystrophies (MD) are an inherited class of disorders characterized by progressive muscle weakness that affects limb, axial, and facial muscles to a variable severity. Fat infiltration of the legs is a clinical manifestation of the disease, which is easily seen in MRI images. MD results in a loss of muscle mass causing a degradation of muscle strength \cite{1}.

Inter-muscular adipose tissue (IMAT) refers to infiltrated fat in the muscle region, and subcutaneous adipose tissue (SAT) refers to the outer fat surrounding the muscle. The two fat tissues are separated by a boundary called "fascia lata" which most of the studies try to detect.

It has been shown that the quantification of fat infiltration based on MRI techniques has a strong correlation with the disease progression and is therefore an accurate marker of disease state and severity \cite{4}. In order to provide physicians with a precise disease bio-marker, an accurate segmentation of muscle tissue, subcutaneous fat and inter-muscular fat is needed.
The "fascia lata" can be obscure and hard to find. 
The most common artifacts challenging the segmentation process are inconsistent pixel intensities and inhomogeneities across the MRI images.

Several works have been published for thigh and calf segmentation. Valentinitsch et al. \cite{7} proposed a three-stage segmentation method using unsupervised multi-parametric $k$-means clustering to segment subcutaneous fat, inter-muscular fat and muscle. Posetano et al. \cite{8} introduced a fuzzy c-means approach, an active contour and Gaussian Mixture Model-Expectation Maximization (GMM-EM) algorithm for subcutaneous fat, muscle, inter-muscular fat and bone segmentation. The main problem facing those approaches is that segmentation using active contour based-methods gives unreliable results when the "fascia lata" is obscure. Chambers et al. \cite{10} introduced muscle-region segmentation method refers to the live-wire approach for path search along the "fascia lata". Tan et al. \cite{11} proposed a deformable model to reconstruct "fascia lata"'s surface. 
With the rapid development of deep-learning and its superior performance, automatic approaches based on Convolutional Neural Networks (CNNs) have been applied recently to the task of IMAT segmentation on thigh and calf MRI. Yao
et al. \cite{13} integrated deep-learning methods with traditional models, proposing a holistic neural networks and dual active contour model for "fascia lata" detection and muscle and IMAT classification. 

In this work, we estimate an index that indicates the stage of the MD disease. This index is the ratio between the IMAT area and the whole region of muscle. Two stages are essential in order to calculate this index accurately. The first stage includes discarding the SAT, bone and bone marrow pixels, leaving us with the muscle-region. The next stage is to discriminate between the healthy muscle pixels and the IMAT pixels within the muscle-region. The "fascia lata" serves the experts a visual separation creating a reliable ground truth (GT) that enables supervised learning methods to accomplish this mission precisely. Thus, encouraged by it's high efficiency in semantic segmentation of small data sets, we employ the U-net architecture for the segmentation of the muscle-region. We show strong performance for a variety of cases, especially for severe fat infiltration, which is the hardest to segment. We also demonstrate the robustness of the network in the existence of MRI artifacts. Following segmentation, pixel-classification is performed to enable the distinction, on a pixel-level, of healthy muscle and IMAT.
Pixels with fat infiltration are usually dispersed over the whole region of muscle. Moreover, the infiltration level varies across pixels. Consequently, the border between healthy muscle and IMAT pixels becomes blurry and a manual pixel-wise labeling becomes difficult and uncertain. Hence, a semi-supervised method in this stage accounts for this limitation. Inspired by \cite{14}, we implement a patch based deep convolutional auto-encoder with a triplet-loss constraint to learn an interpretable latent feature representation and apply $k$-means in the embedded space to classify the pixels into two clusters. The integration of patches helps tackle the problem of small data, exploits the contextual information of the pixels and maintains the relationship to adjacent pixels. Results demonstrate the relevance of clustering to our task, and the ability of the overall system to predict the level of fat infiltration.  

\begin{figure}[t]
\centering
\includegraphics[width=10cm]{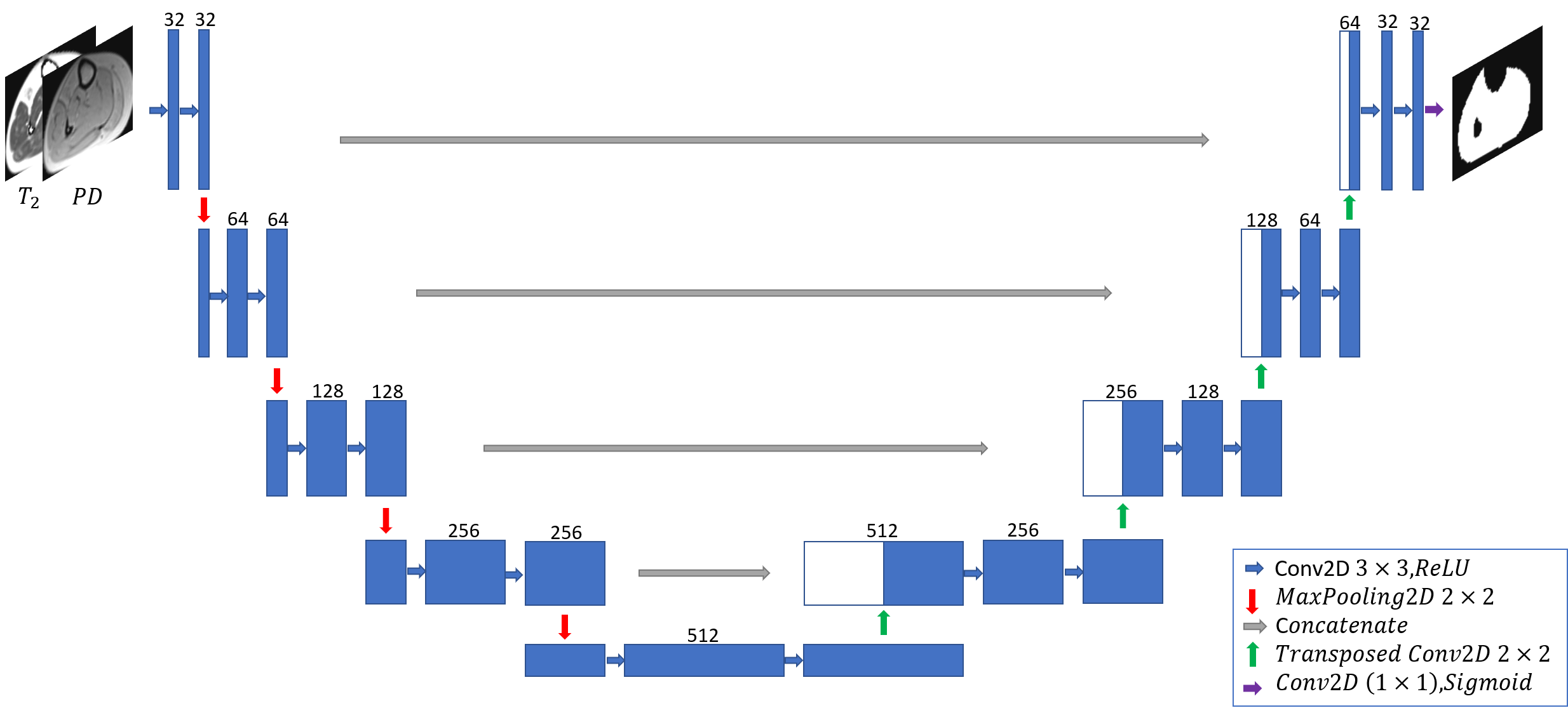}
\caption{The architecture of U-net used for muscle-region segmentation.}
\label{fig:1}
\end{figure}
\section{Methods}

\subsection{Data set}
Our data set includes 17 axial MR scans of patients' legs (thigh and calf) suffering from muscular disorders. A GT of the muscle regions for all the images was delineated by a domain expert.

The MR scans were scanned on a whole-body Siemens Prisma 3T scanner. A multi spin-echo sequence was used with relaxation time $(TR) = 1479\ ms$, echo time $(TE) = 8.7\ ms$, echo train length $(N_{echo}) = 17$, spatial resolution $ = 1.5\times1.5\ mm^2$ with a matrix size of $128\times128$. The regions were imaged acquiring 5 slices with slice thickness of $10\ mm$, the acquisition time for one scan is $5:07\ minutes$. 

To construct the $T_2/PD$ maps, Bloch simulations were used to estimate the actual echo modulation curve (EMC). Simulations were repeated for a range of $T_2$, transmit field ($B_1^+$) and proton density ($PD$) values yielding a database of simulated EMCs, until the [$T_2$, $PD$, $B_1^+$] set of values mostly matched the measured data at each voxel \cite{13a}.

IMAT and healthy muscle ground truth is calculated on a pixel-by-pixel basis. The infiltration of subcutaneous fat into the diseased muscle region causes a mixture of two $T_2$ components to appear in each imaged voxel. An extension of the EMC algorithm was introduced \cite{13e} which is based on a two-$T_2$ component decomposition of the signal in each imaged voxel, simultaneously estimating fat and water fractions within a single voxel. Voxels whose fat fraction was $> 50\%$ were labelled as fat (i.e., diseased muscle), and the rest were labelled as muscle (i.e., healthy muscle). The fraction of IMAT area relative to the whole muscle region was calculated.
\begin{figure}[t]
\centering
\includegraphics[width=12cm]{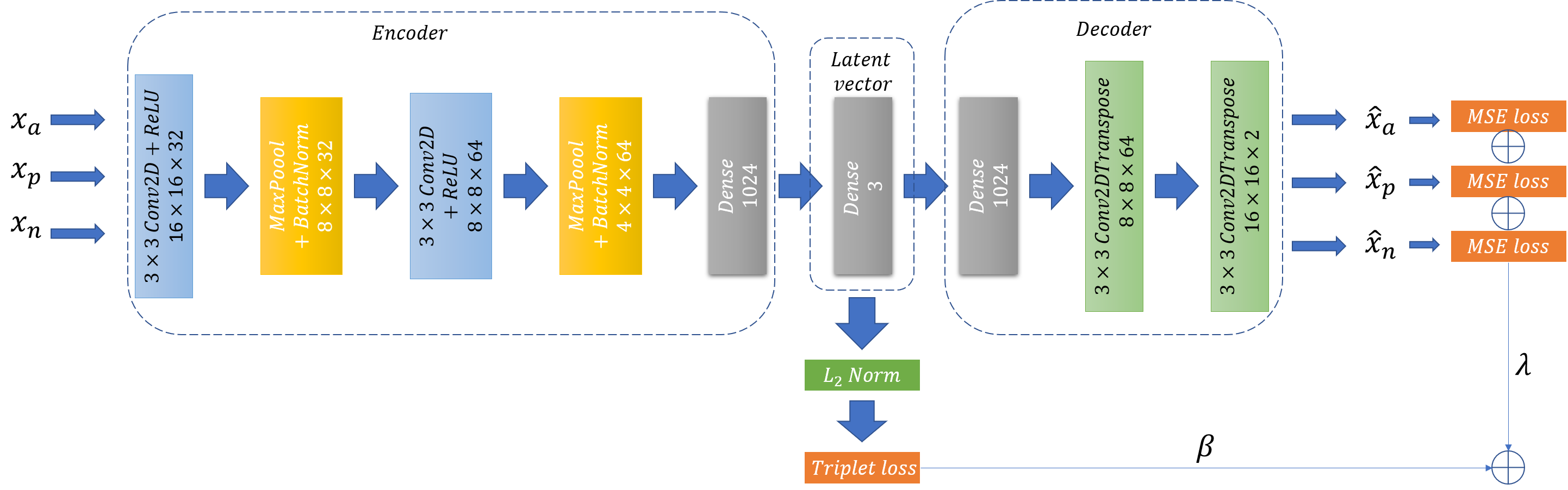}
\caption{The architecture of the deep convolutional auto-encoder and the losses used for muscle tissue classification.}
\label{fig:2}
\end{figure}
\subsection{Muscle-Region Semantic Segmentation}
In the first stage, we train a neural network to segment the region inside the "fascia lata". The subcutaneous adipose tissue (SAT), bone and bone marrow are masked out simultaneously by this method.

In the preprocessing stage, we separate the thighs$\setminus$calf regions from the background by applying a canny edge detector in order to detect the outer edge. Then, we crop the image around the region of interest. The cropped image is then resized to ($128\times128$).
The intensity inhomogeneity that is inherent to the MRI images is corrected by using the N4ITK method \cite{15}. 
After producing the $T_2$ and $PD$ maps, we correct for extreme values in individual pixels within the $T_2$ images by taking the $98\%$ percentile of the image intensity range and clipping the pixels above this value.
We normalize each image to zero mean and unit variance before proceeding to the next step.

We employed a popular fully convolutional network (FCN)-based deep learning architecture, U-net \cite{17}, for the segmentation of muscle region. This U-net network has been demonstrated to work well on medical images with a very few learning samples and a strong use of data augmentation. The network architecture is illustrated in Fig.~\ref{fig:1}, The left part of the network is a contracting path and the right part is a symmetrical expanding path which decompresses the features back to its original size. The concatenating path consists of five levels with different resolution feature maps. Each level consists of two layers of $3\times3$ non-padded convolutions followed by a rectified linear unit (ReLU). Following the two convolution layers, we apply a $2\times2$ max pooling operation with stride $2$ for down-sampling. After each down-sampling step we double the number of feature channels in the next two convolution layers.
The expansive path consists of five levels, where in each we halve the feature channels number. In each level we use a transposed convolution (a.k.a deconvolution) with a $2\times2$ kernel size and stride of 2. The transposed convolution optimally learns the up-sampling, which helps restore the image more precisely than using interpolation for up-sampling. Next, we concatenate those maps with the corresponding feature maps from the contracting path and apply two $3\times3$ convolutions, each followed by a ReLU. At the final layer a $1\times1$ convolution is used to map the feature vector to the desired number of classes.

Four types of inputs to the FCN were considered in this work: Two FCNs were fed by a concatenated $T_2$ and $PD$ maps (2 channels). The difference between the two inputs lies in the preprocessing stage: one input was fed as it is without preprocessing, while the second input's inhomogeneous intensity was corrected and the $T_2$ maps were clipped. Two additional FCNs were used: in one network $T_2$ maps were the input (raw and preprocessed); in the other, $PD$ maps were the input (raw and preprocessed).  
The purpose of these experiments was to explore the FCN's ability to reach excellent results in the existence of MRI artifacts and in the absence of inhomogeniety correction methods, as was used in previous works. 
The training set included 14 patients with the corresponding binary segmentation maps. The models were trained using Adam optimizer, with the default settings: $lr=0.001$, $\beta_1=0.9$, $\beta_2=0.999$ and $\epsilon=10^{-8}$. The loss that was optimized is soft Dice coefficient loss. The batch size was set to 8 and the models were trained for 100 epochs. We found it necessary to augment the original training images, increasing the number of images by 10, in order to improve the model's robustness in the presence of data variance. We randomly used shift (0.2 of image height and width), zoom (between 0.9 to 1.3 of image size), rotation ($0^\circ-30^\circ$) and flip (vertical$\setminus$horizontal).
\begin{figure}[t]
\centering
\includegraphics[width=9cm]{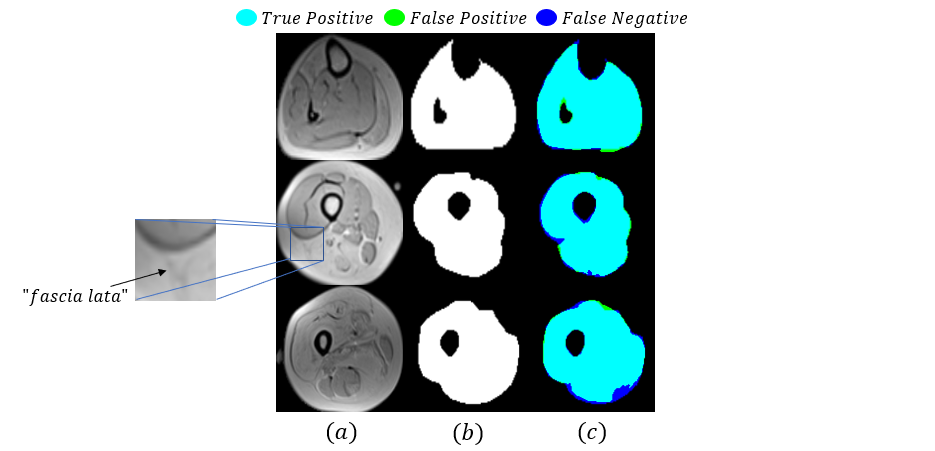}
\caption{\textbf{Examples of muscle segmentation:} (a) MRI images of three different patients: from top to bottom: mild, moderate and severe fat infiltration, (b) Delineated GT of muscle region used for training and testing, and (c) Overlap between GT and the output of the FCN.}
\label{fig:3}
\end{figure}
\subsection{Healthy Muscle and IMAT Classification}
We further classified the region of muscle that we have segmented into two types of tissue: healthy muscle and IMAT pixels. We employed a patch-based deep convolutional auto-encoder (DCAE) to learn semantic feature representation incorporating deep metric learning. The learned embedding is utilized for tissue clustering with $k$-means algorithm. This method is motivated by our need to learn embedded feature representation of the two tissues with a constraint that imposes similarity across the patches of the same tissue and non-similarity between patches of different tissues.  
\subsubsection{Deep Convolutional Auto-encoder and Triplet Loss (DCAE$_{TL}$).}
Since the clustering is performed for each pixel, we trained the DCAE on patches of size $16\times16\times2$ extracted around each pixel from the $T_2$ and $PD$ images of the muscle region. The DCAE is composed of an encoder and a decoder. The encoder consists of two blocks with 32 and 64 feature maps. Each block is built of $3\times3$ convolutions followed by ReLU, $2\times2$ max pooling and batch-normalization. The output of the two fully convolution blocks is then flattened into 1024 units and followed by a dense layer that encodes the features in the embedded space. This is followed by $L_2$ normalization that constrains the embedding to live on a hypersphere. The decoder structure is composed of dense layer of 1024 reshaped to size $4\times4\times64$, followed by $3\times3$ transposed convolution layer with 32 filters and stride 2, and another $3\times3$ transposed convolution layer that reconstructs the input patch (see Fig.~\ref{fig:2}). The DCAE was trained with Adam optimizer with default settings, and batch size of 256 for 100 epochs.
\begin{table}[t]
\centering
\caption{Quantitative comparison between the performance of the FCN on different input data types for muscle region segmentation (pp stands for preprocessing).}
\begin{tabular}{|c|c|c|c|c|c|}
    \hline
    \multirow{2}{*}{Input} &
        \multicolumn{4}{c|}{DSC} \\
    & Mild & Moderate & Severe & Combined \\
    \hline
    $T_2$+$PD$ with pp & 0.971 & 0.926 & \textbf{0.964}  & 0.956 \\
    \hline
    $T_2$+$PD$ w/o pp & 0.973 & 0.959 & 0.949 & 0.958 \\
    \hline
    $T_2$ with pp & 0.962 & 0.883 & 0.940  & 0.931 \\
    \hline
    $T_2$ w/o pp & 0.958 & 0.959 & 0.938  & 0.948 \\
    \hline
    $PD$ with pp & \textbf{0.974} & \textbf{0.962} & 0.962 & \textbf{0.964} \\
    \hline
    $PD$ w/o pp & 0.970 & 0.960 & 0.960 & 0.962 \\
    \hline
\end{tabular}
\label{table:1}
\end{table}
\begin{table}[t]
\centering
\caption{Performance assessment of muscle tissue classification of our method compared to other clustering methods.}
\begin{tabular}{|c|c|c|c|c|c|}
    \hline
    Method & Healthy Muscle Dice & IMAT Dice & ACC & NMI & ARI \\ 
    \hline
    $k$-means & 0.877 & 0.906 & 0.906  & 0.565 & 0.654\\ 
    \hline
    DCAE + $k$-means  & 0.895 & 0.915 & 0.917 & 0.586 & 0.692\\ 
    \hline
    DCAE$_{TL}$ + $k$-means & \textbf{0.911} & \textbf{0.933} & \textbf{0.933} & \textbf{0.627} & \textbf{0.746}\\ %
    \hline
\end{tabular}
\label{table:2}
\end{table}
The loss consists of two parts: reconstruction loss and triplet loss. The reconstruction loss is the mean squared error (MSE). To train, we randomly select input patch triplet $(x_a^i, x_p^i, x_n^i)$ from the training set. The distance of the anchor patch ($x_a^i$) from the positive patch ($x_p^i$), that roughly matches the anchor patch and the center pixel has same label as the anchor, is smaller than the distance from the negative patch ($x_n^i$). The auto-encoder is trained simultaneously on the three patches, transforms them to latent vectors $f(x_a^i)$, $f(x_p^i)$ and $f(x_n^i)$ and reconstructs each one to the original image $(\hat{x}_a^i, \hat{x}_p^i, \hat{x}_n^i)$. The latent vectors are used for the calculation of the triplet loss ($L_{triplet}$). The triplet loss over a batch $N$ can be expressed as follows:
\noindent
\begin{equation} 
\label{eq:1}
L_{triplet} = \sum_{i=1}^{N}\max{\Big\{0, ||f(x_a^i) - f(x_p^i)||_2^2 - ||f(x_a^i) - f(x_n^i)||_2^2 + \alpha \Big\}}
\end{equation}
where $\alpha$ is a margin that is enforced between positive and negative pairs and set to 1 in our experiments. The combined loss is defined in equation \ref{eq:2}:
\noindent
\begin{equation} 
\label{eq:2}
L_{total} = \beta L_{triplet} + \lambda(L_{MSE}(x_a^i, \hat{x}_a^i) + L_{MSE}(x_p^i, \hat{x}_p^i) + L_{MSE}(x_n^i, \hat{x}_n^i))
\end{equation}
Where $\beta$ and $\lambda$ are loss weights, experimentally set to $1/2$ and $1/6$, respectively.
We compared this method with the classical $k$-means applied to the $T_2$ and $PD$ values of each pixel and with deep convolutional auto-encoder followed by $k$-means (DCAE + $k$-means), where the DCAE is trained over patches with MSE loss and the $k$-means is applied to the embedded layer.

\begin{figure}[b]
\centering
\includegraphics[width=6cm]{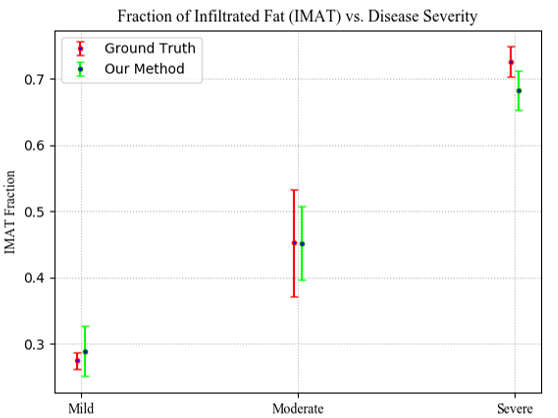}
\caption{Infiltrated fat (IMAT) as a fraction of the whole muscle in mild, moderate and severe cases compared to the GT.}
\label{fig:4}
\end{figure}
\section{Results}
The test set includes three patients with mild, moderate and severe fat infiltration.
We evaluated the approach performance in terms of Dice Similarity Coefficient (DSC) of the predicted delineation to the GT annotation. The clustering performance of the evaluated clustering methods was evaluated with respect to the normalized mutual information (NMI), accuracy of clustering (ACC), and adjusted Rand index (ARI). NMI is a similarity measurement borrowed from information theory based on the mutual information of the ground-truth classes and the obtained clusters and normalized using the entropy of each. ARI is a variant Rand index that is adjusted for the chance grouping of elements. We also calculated the DSC for both healthy muscle and IMAT. 
The results for the muscle region segmentation are presented in Table~\ref{table:1} and show very high performance for the variety of inputs proving reliability and robustness of the method. 
Fig.~\ref{fig:3} shows results for muscle region segmentation. Our method succeeds to segment the muscle region with a small number of false positives and negatives around the "fascia lata" even when it becomes obscure and blurry in moderate and severe cases where the fat infiltration is high. Table~\ref{table:2} shows the results of clustering for the three clustering methods. The proposed method outperforms the competing methods in healthy muscle dice, IMAT dice, accuracy, NMI and ARI. Fig.~\ref{fig:4} shows the fraction of infiltrated fat area from the whole muscle region (healthy muscle + IMAT) which indicates the severity of fat infiltration. In addition to being agreeable with the corresponding GT, the fractions show a high correlation between our calculated index and the severity of the disease.

\section{Conclusions}
This work presents a robust method to segment the muscle-region and classify the pixels within the region to healthy muscle pixels and IMAT pixels. We demonstrated the reliability of the proposed method when intensity inhomogeneity artifacts exist in MRI images. We tested the method on patients with mild, moderate and severe muscular dystrophies and proved that the method overcomes the issue of identifying the borders between tissues when the disease is in an advanced stage. We also calculated the fraction of the infiltrated fat and showed a high correlation between this index and the disease severity.  


\end{document}